\documentclass[preprint,prl,aps,superscriptaddress,nopacs]{revtex4}

\usepackage{graphicx}
\usepackage{color}

\begin{document}

\title{From Anomalous Hall Effect to the Quantum Anomalous Hall Effect}

\author{Hongming Weng}
\author{Xi Dai}
\author{Zhong Fang}  % \email{zfang@iphy.ac.cn}
\affiliation{ Beijing National Laboratory for Condensed Matter
  Physics, and Institute of Physics, Chinese Academy of Sciences,
  Beijing 100190, China } \date{Aug. 24, 2013}

\begin{abstract}
A short review paper for the quantum anomalous Hall effect. A substantially extended one is published as {\it Adv. Phys.} {\bf 64}, 227 (2015).
\end{abstract}

%\pacs{74.70.-b, 74.25.Jb, 74.25.Ha, 71.20.-b}
\maketitle

%
% Historical review of family of hall effect
% 

In 1879, Edwin H. Hall discovered that when a conductor carrying
longitudinal current was placed in a vertical magnetic field, the
carrier would be pressed to against the transverse side of the
conductor, which led to the observed transverse voltage. This is
called Hall effect (HE)~\cite{halleffect}, which is a remarkable
discovery, in spite of that the understanding of it was difficult at
that time since electron was to be discovered 18 years later. After
that, as we know, HE is well understood and it is the Lorentz force
experienced by the moving electrons in magnetic field pressing them
against the transverse side. The HE is now widely used as to measure
the carrier density or the strength of magnetic fields.

In 1880, Edwin H. Hall further found that this "pressing electricity"
effect can be larger in ferromagnetic (FM) iron than in non-magnetic
conductors.  The enhanced Hall effect comes from the additional
contribution of the spontaneous long range magnetic ordering, which
can be observed even without applying external magnetic field. To be
distinguished from the former one, this is termed as anomalous Hall
effect (AHE)~\cite{ahe}. Though HE and AHE is quit similar to each
other, the underlying physics are much different since there is no
orbital effect (Lorentz force) when external magnetic field is absent
in AHE.

The mechanism of AHE has been a enigmatic problem since its discovery,
and it lasted for almost a century. The AHE problem involves concepts
deeply related with topology and geometry that have been formulated
only in recent years after the Berry phase being recognized in
1984~\cite{berry}. In hindsight, Karplus and Luttinger~\cite{kl}
provided a curtail step in unraveling this problem as early as in
1954. They showed that moving electrons can have an additional
contribution to its group velocity when an external electric filed is
applied. This additional term, dubbed "anomalous velocity", which is
contributed by all occupied band states in FM conductors with
spin-orbit coupling (SOC), can be non-zero and leads to the
AHE. Therefore, this contribution depends only on the band structure
of perfect periodic Hamiltonian and is completely independent of
scattering from impurities or defects (therefore called intrinsic
AHE), which makes it hard to be widely accepted before the concept of
Berry phase being well established. In a long time, two other
"extrinsic" contributions had been thought as the dominant mechanisms
that give rise to an AHE. One is the skew scattering~\cite{skew} from
impurities caused by effective SOC and the other one is the side
jump~\cite{side} of carriers due to different electric field
experienced when approaching and leaving an impurity. The controversy
arises also because it is hard to make quantitive comparison with
experimental measurements. The unavoidable defects or domains in
samples are complex and hard to be treated quantitively within any
extrinsic model and the contributions from both "intrinsic" and
"extrinsic" mechanisms co-exist.

In 1980, K. von Klitzing {\it et al.} made remarkable discovery of
quantum Hall effect (QHE)~\cite{qhe}. He found that, with the increase
of external magnetic filed, the Hall conductivity exhibits a serials
of quantized plateau, and at the same time, the longitudinal
conductivity becomes zero, i. e., the sample bulk becomes insulating
when Hall conductivity is quantized.  In the QHE, electrons
constrained in two-dimensional (2D) samples, are enforced to change
its quantum states into new ones, namely the Landau energy levels,
under highly intensive magnetic field. The originally free-like
conducting electrons start to make cyclotron motion. If the magnetic
field is strong enough, such cyclotron motions will form full circles,
which makes electrons localized in the bulk and the sample becomes
insulating. While along the edges of the 2D system, the circular
motion enforced by the magnetic field can not be completed due to the
presence of the edges, which enforce the electrons to travel in one
way forming so called edge states. The electrons in such state can
circumambulate defects or impurities on their way
"smartly". Therefore, current carried by these electrons is
dissipationless and conductance is quantized into unit of $e^2/\hbar$
with quantum number corresponding to the number of edge states. Such
fascinating quantum state and physical phenomena are highly
interesting and impact the whole field of physics, because this is a
new state of condensed matters and it should be characterized by the
topology of electronic wave-function~\cite{laughlin}. This topological
number is given by D. J. Thouless, M. Kohmoto, M. P. Nightingale and
M. den Nijs~\cite{tknn} in 1982 and is called as the TKNN number or
the first Chern number, which has direct physical meaning as the
number of edge states or the quantum number of Hall conductance.

After reaching this point, one may immediately ask the question: Can
we have the quantum version of AHE, similar to the QHE?
Unfortunately, this question was irrelevant at that historical moment,
because we even did not know yet the fundamental mechanism of AHE
then. Nevertheless, in 1988, F. D. M. Haldane proposed~\cite{haldane}
that a QHE without any external magnetic field is in principle
possible. Although this proposal is very simple and has nothing to do
with either the AHE or the SOC, his idea play important roles for many
of our nowadays studies, such as the topological insulators and the
quantum anomalous Hall effect (QAHE), because he pointed out the
possibility to have a novel electronic band structure of perfect
crystal, which carry non-zero Chern number even in the absence of
external magnetic field.

\begin{figure}[tbp]
\includegraphics[clip,scale=0.3]{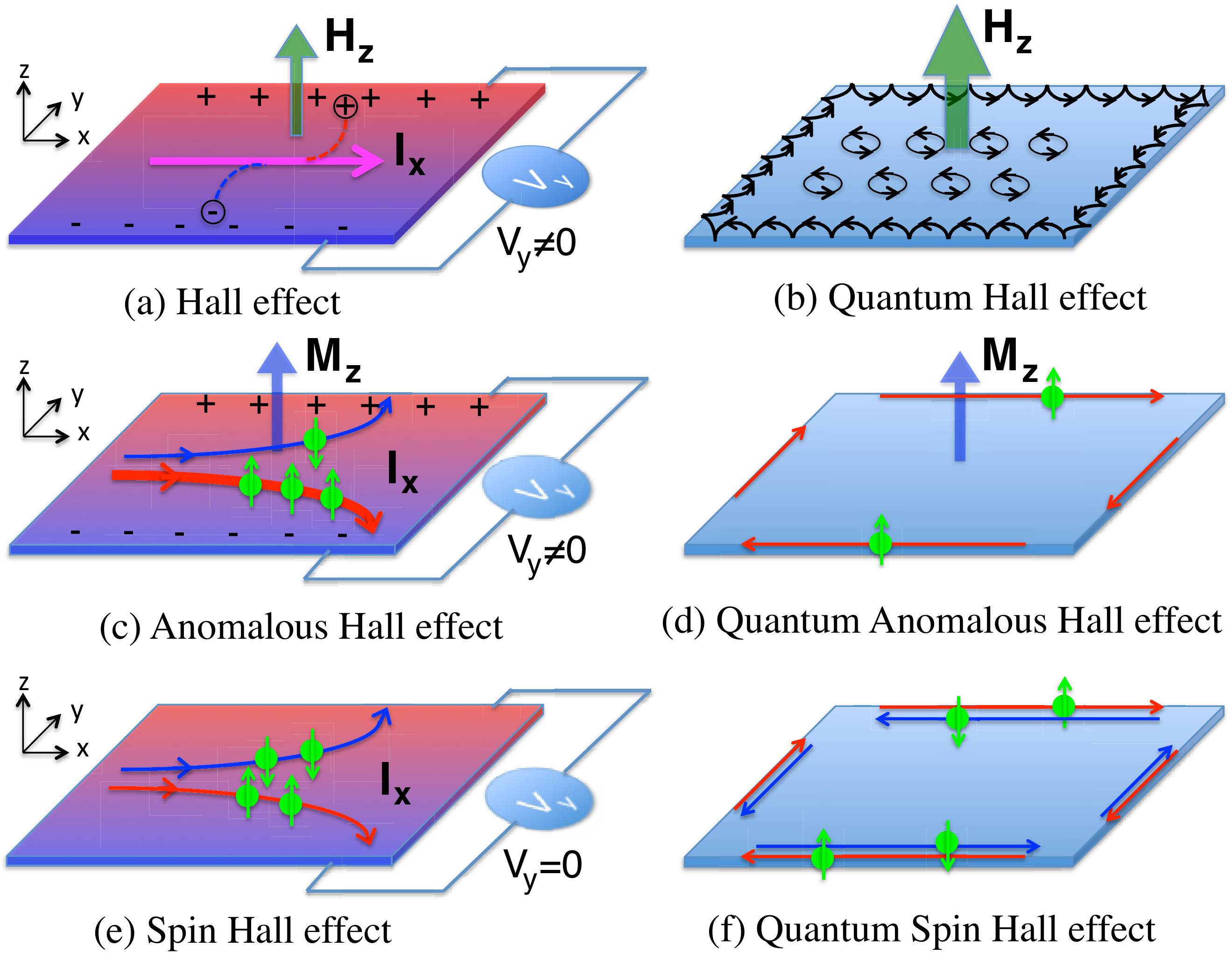}
\caption{(Color online) (a) Hall effect. The longitudinal current
  I$_x$ under vertical external magnetic filed H$_z$ contributes to
  the transversal voltage V$_y$ due to the Lorentz force experienced
  by carriers.  (b) Quantum Hall effect. The strong magnetic field
  H$_z$ enforces electrons into Landau level with cyclotron motion and
  become localized in bulk while conducting at edges. (c) Anomalous
  Hall effect. The electrons with majority and minority spin (due to
  spontaneous magnetization M$_z$) having opposite "anomalous
  velocity" due to spin-orbit coupling, which causes unbalanced
  electron concentration at two transversal sides and leads to finite
  voltage V$_y$. (d) Quantum Anomalous Hall effect.  The nontrivial
  quantum state satisfies all necessary conditions and leads to
  insulating bulk while topologically protected conducting edge state
  with spontaneous magnetization. (e) Spin Hall effect. In nonmagnetic
  conductor, equivalent currents in both spin channels with opposite
  "anomalous velocity" leads to balanced electron concentration at
  both sides while net spin current in transversal direction. (f)
  Quantum Spin Hall effect.  The 2D nontrivial Z$_2$ insulator has
  conducting edge states with opposite spin in different direction,
  which can be viewed as two time reversal symmetrical copies of QAHE.}
\end{figure}

The underlying physics of AHE, in particular, the topological nature
of intrinsic AHE was not fully appreciated only until the early years
of 21 century. In a series of papers, (i.e, Jungwirth, et.al
(2002)~\cite{jungwirth}; Onoda, et.al. (2002)~\cite{onoda}; Fang,
et.al (2003)~\cite{fang}; Yao, et.al. (2004)~\cite{yao}), it was
discovered that the intrinsic AHE can be related to the Berry phase of
the occupied Bloch states. The so-called "anomalous velocity'' is
originated from the Berry-phase curvature, which can be regarded as
effectively magnetic field in the momentum space, and thus modifies
the equation of motion of electrons, leading to the AHE.  This
effective magnetic field can be also traced back to the band crossings
and the magnetic monopoles in the momentum space~\cite{fang}, which is
now called Weyl nodes in the Weyl semimetals~\cite{weyl,HgCrSe}. The
quantitative first-principles calculations for SrRuO$_3$~\cite{fang}
and Fe~\cite{yao}, in comparison with experiments, suggested that
intrinsic AHE actually dominates over extrinsic ones. The presence of
SOC and the breaking of time reversal symmetry (due to the FM
ordering) are crucially important, because otherwise the Berry phase
contribution will be prohibited by symmetry. What makes those
understandings unique is that, like the QHE, the intrinsic AHE is now
directly linked to the topological properties of the Bloch states. Up
to this point, it is now proper to ask the question: Can the AHE also
be quantized like its cousin HE? If it is realized, we will have a
kind of QHE without magnetic field, although it is already much
different from Haldane's original speculation in the sense that SOC
must play important roles here. Nevertheless, from the applicational
point of view, the realization of QAHE will certainly stimulate the
wide usage of such novel quantum phenomena in future technology, in
particular, the dissipationless edge transport without magnetic field.

To reach this goal, there is still a big step to be overcome. The QAHE
requires four necessary conditions: (1) the system must be 2D; (2)
Insulating bulk; (3) FM ordering; and (4) non-zero Chern number. It
may be easy to satisfy one or two conditions, but hard to realize all
of them simultaneously. Fortunately, the recently rapid progresses in
the studies of topological insulators (TI)~\cite{topo1,topo2} make the
challenge of realizing QAHE possible. The TI is a new state of quantum
matter, which is characterized by the $Z_2$ topological
number~\cite{kanemele} and is protected by the time-reversal symmetry
(TRS), in contrast to the QHE and QAHE, where the TRS must be
broken. However, an important view is that the 2D topological
insulator and the quantum spin Hall effect (QSHE)~\cite{qshe,hgte} can
be effectively viewed as two copies of distinct QAHE states which are
related by the TRS, in other words the 2D QSHE state can be viewed as
the time-reversal invariant version of QAHE state. Given the known
material realizations of 2D and 3D
TI~\cite{hgte,hgteexp,topo1,Zhang,Xia,Chen}, it is natural to start
from those systems and try to break the TRS in order to achieve the
QAHE. Following this strategy, several possibilities are proposed
theoretically. Qi, et.al.~\cite{qi} first pointed out that gapping the
Dirac type surface states of 3D TI by FM insulating cap-layers may
produce the QAHE, although their arguments are not concrete. Later on,
the "band inversion" picture and the experimental observation of
QSHE~\cite{hgte,hgteexp} inspired the idea of realizing QAHE by
transition metal (Mn) doped HgTe quantum well
structure~\cite{liucxMnHgTe}. Unfortunately, in such case, the Mn
local moments do not order ferromagnetically. In 2010, Yu,
et.al,~\cite{yuqahe} predicted that when a topological insulator
Bi$_2$Se$_3$ or Bi$_2$Te$_3$ is made thin and magnetically doped (by
Cr or Fe), the system should order ferromagnetically through the van
Vleck mechanism, and exhibit the QAHE with a quantized Hall resistance
of $h/e^2$ --- a proposal that was finally achieved experimentally by
Chang, et.al.~\cite{qaheexp1,qaheexp} after great efforts.

While this is not the end of the story, instead, we believe this
success will inspire more extensive researches on QAHE. Two important
issues become the natural directions for the future studies: (1) how
to increase the temperature range of QAHE (it is now observed only in
the tens mK range); (2) how to realize the higher plateau with Chern
number larger than 1.  There are several other proposals worth trying.
HgCr$_2$Se$_4$ has been predicted to be a Weyl semimetal, and its
quantum well structure with proper thickness would exhibit QAHE with
Chern number being 2~\cite{HgCrSe}, different from the Cr-doped
Bi$_2$Te$_3$ family thin film. The advantage of this proposal is that
HgCr$_2$Se$_4$ is a chemically pure and stable compound with known
bulk Curie temperature higher than 100K.  One similar proposal is
GdBiTe$_3$~\cite{gdbite}, the thin film of which have one edge state
contributing to QAHE. Garrity and Vanderbilt~\cite{garrity} proposed
that Au, Pb, Bi, Tl, I, and etc heavy-element layers on the surface of
magnetic insulators may contribute to large AHE and even QAHE. With
the further material breakthroughs, we have strong reason to expect
that the QAHE may someday find its places in our electronic devices.

This work was supported by the National Science Foundation of China
and by the 973 program of China.

\end{document}